\documentclass[conference,compsocconf]{IEEEtran}

\usepackage{epsfig}
\usepackage{graphicx}
\usepackage{caption}
\usepackage{amssymb}

\usepackage{algorithm}
\usepackage[noend]{algpseudocode}
\usepackage{multirow}
\usepackage{balance}
\usepackage{sidecap}

\usepackage[caption=false]{subfig}

\usepackage{color,soul}
\usepackage{xcolor}

\usepackage{amsthm}

\usepackage{color}
\usepackage{listings}
\begin{document}

%

\title{Run-time Parameter Sensitivity Analysis Optimizations}

\author{Eduardo Scartezini$^1$, Willian Barreiros Jr$^1$., Tahsin Kurc$^{2,3}$,\\ Jun Kong$^4$, Alba C. M. A. Melo$^1$, Joel Saltz$^2$, and George Teodoro$^{2,5}$,\\
\IEEEauthorblockA{$^1$Department of Computer Science, University of Bras\'ilia, Bras\'ilia, DF, Brazil\\
$^2$Department of Biomedical Informatics, Stony Brook University, Stony Brook, NY, USA \\
$^3$Scientific Data Group, Oak Ridge National Laboratory, Oak Ridge, TN, USA\\
$^4$Biomedical Informatics Department, Emory University, Atlanta, GA, USA\\
$^4$Department of Computer Science, Universidade Federal de Minas Gerais, Belo Horizonte, MG, Brazil\\
wbarreiros@aluno.unb.br,\{albamm\}@unb.br,jun.kong@emory.edu,\{tahsin.kurc,joel.saltz\}@stonybrook.edu,george@dcc.ufmg.br}}

\maketitle
\vspace*{-2ex}
\begin{abstract}

Efficient execution of parameter sensitivity analysis (SA) is critical to allow
	for its routinely use. The pathology image processing application
	investigated in this work processes high-resolution whole-slide cancer
	tissue images from large datasets to characterize and classify the
	disease. However, the application is parameterized and changes in
	parameter values may significantly affect its results. Thus,
	understanding the impact of parameters to the output using SA is
	important to draw reliable scientific conclusions. The execution of the application is rather compute intensive,
	and a SA requires it to process the input data multiple times as
	parameter values are systematically varied. Optimizing this process is
	then important to allow for SA to be executed with large datasets. In
	this work, we employ a distributed computing system with novel
	computation reuse optimizations to accelerate SA. The new computation
	reuse strategy can maximize reuse even with limited memory availability
	where previous approaches would not be able to fully take advantage of
	reuse. The proposed solution was evaluated on an environment with 256
	nodes (7168 CPU-cores) attaining a parallel efficiency of over 92\%,
	and improving the previous reuse strategies in up to 2.8$\times$. 

\end{abstract}
%
%
%
%
\begin{IEEEkeywords} Microscopy Imaging; Sensitivity Studies; Memory-Aware Scheduling.
\end{IEEEkeywords}
\section{Introduction} \label{sec:intro}

The process of quantifying the impact of the input parameters of an application
workflow on its outputs is defined as sensitivity analysis
(SA)~\cite{saltelli2004sensitivity}. This analysis is carried out by
re-executing the target application and quantifying the output results changes
as parameters' values are modified. The use of SA methods is important as it
can~(i) improve our understanding on the correlation between workflows' inputs
and outputs, (ii)~improve the quality/stability of the application workflow
output by identifying sources of uncertainty, and (iii)~simplify workflows by
either fixing parameters or removing parts with little impact on the output.

\begin{figure}[htb!]
\begin{center}
\includegraphics[width=0.47\textwidth]{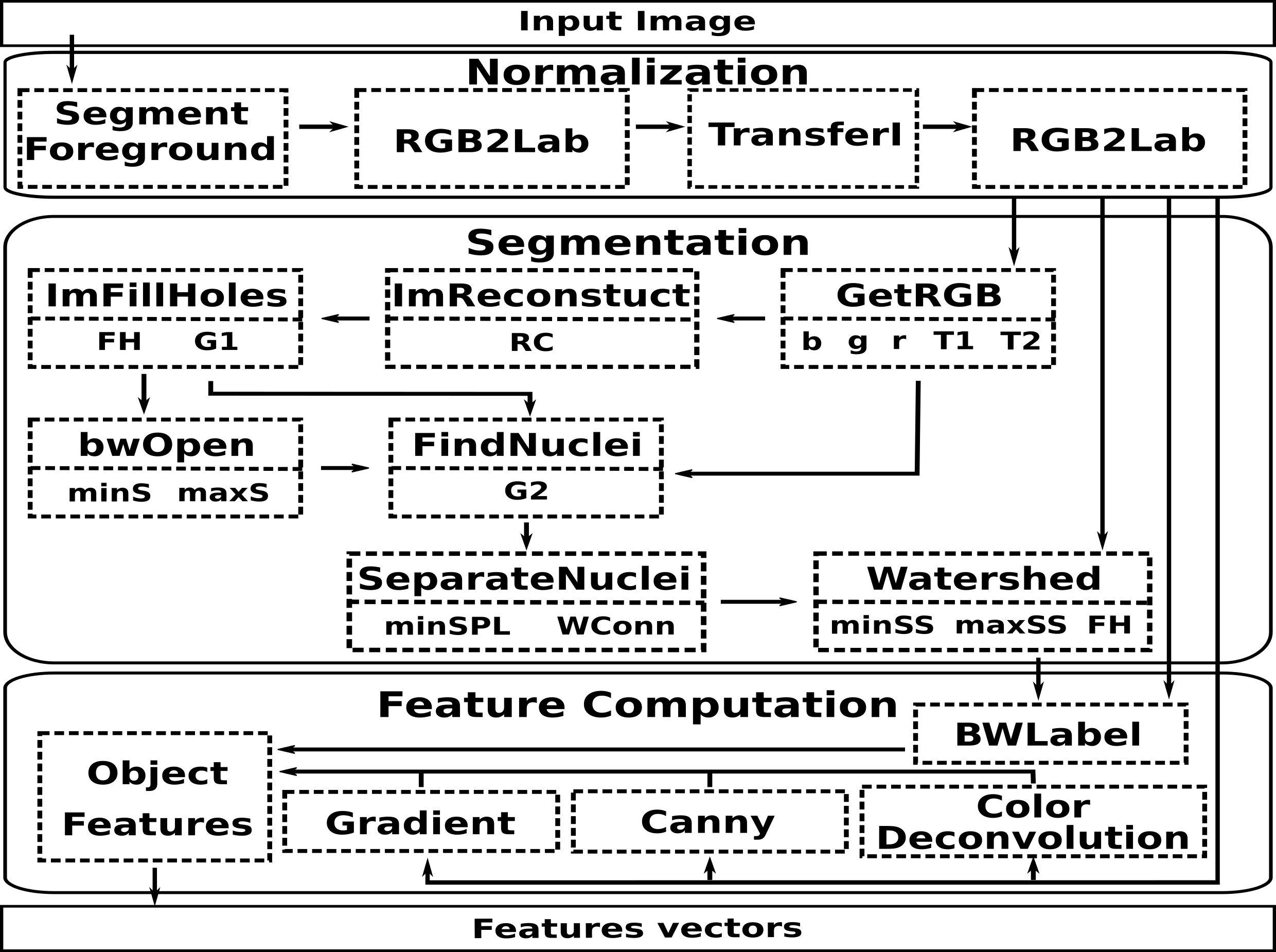}
\vspace*{-1ex}
\caption{Motivating application stages: normalization, segmentation and feature computation stages. Operations used in each stage and parameters are shown with parameters.}
\vspace*{-3ex}
\label{fig:example-app}
\end{center}
\end{figure}

The pathology image analysis domain that motivates this work may benefit from
SA. A typical application in this area processes whole-slide tissue images
(WSI) that may have in the order of 120K$\times$120K pixels or 50GB in size
through a series of transformations that include normalization, segmentation,
feature computations, and other correlative analysis. The first three stages are
more compute intensive and output segmented objects (e.g., cells'
nuclei) along with their shape and texture data features. This information may
be used in several ways, for instance, to perform survival
correlations. Our motivating application workflow is presented in
Figure~\ref{fig:example-app}, which shows the main computing stages and their
internal workflow of tasks.

Much work has been done to adapt  and employ SA methods for other
domains~\cite{10.2307/1269043,Weirs2012157,Campolongo20071509,Bertrand-2015}.
Nevertheless, the use of SA in practice can be challenging given that the many
applications are very compute demanding. This is the case in pathology
image analysis, where the execution of a single WSI can take hours when
processed in a node. This is worsen by the fact that an analysis study will
employ hundreds of WSIs and our application has several parameters
(Table~\ref{tab:parameters}). Thus, processing all images as
parameters are changed in a SA is a very costly process, which motives this
work.

\begin{table}[h]
\begin{center}
\vspace*{-2ex}
\caption{Application parameters and their range values~\cite{8048914}. }
\begin{scriptsize}
\begin{tabular}{l l l l }
\hline
Parameter		& Description   					& Range Values			\\ \hline
B/G/R  			& Background detection thresholds 			&  $[210,220,...,240]$ 	\\ \hline
T1/T2			& Red blood cell thresholds				& $[2.5,3.0,...,7.5]$		\\ \hline
\multirow{2}{*}{G1/G2}	& Thresholds to identify 				& $[5,10,...,80]$		\\ 
			& candidate nuclei					& $[2,4,...,40]$		\\ \hline
MinSize(minS)		& Candidate nuclei area threshold   			& $[2,4,...,40]$ 		\\ \hline
MaxSize(maxS)		& Candidate nuclei area threshold 			& $[900,..,1500]$		\\ \hline 
MinSizePl \\ (minSPL)	& Area threshold before watershed  			& $[5,10,...,80]$		\\ \hline
MinSizeSeg \\ (maxSS)	& Area threshold in final output			& $[2,4,...,40]$ 		\\ \hline 
MaxSizeSeg \\ (minSS)	& Area threshold in final output  			& $[900,..,1500]$		\\ \hline 
FillHoles(FH) 		& propagation neighborhood				& $[4$-conn$,8$-conn$]$		\\ \hline
MorphRecon(RC)		& propagation neighborhood				& $[4$-conn$,8$-conn$]$		\\ \hline
Watershed(WConn)	& propagation neighborhood				& $[4$-conn$,8$-conn$]$		\\ \hline
\end{tabular}
\end{scriptsize}
\label{tab:parameters}
\vspace*{-4ex}
\end{center}
\end{table}

In a previous work, computation reuse has been  evaluated as an optimization to
speedup SA~\cite{8048914,george2016}. Computation reuse opportunities occur in
SA studies because the same dataset is processed multiple times as parameters
sets are systematically varied. These parameter sets may have subsets of
parameters with the same values, which allows for parts of the application
workflow to be reused. In our example application
(Figure~\ref{fig:example-app}), there is opportunity for reuse either of entire
stage instances (coarse-grain) stages when parameters used by instances are the
same, or among a subset of the internal tasks (fine-grain) when only a parts of
the parameters are equal. A previous work~\cite{8048914} has proposed a
strategy called Reuse-Tree Merging Algorithm (RTMA) that may take advantage of
reuse in both granularities. However, its reuse capabilities are limited
because the memory required by a stage to executed increases proportionally
with the computation reuse employed. This problem is worsened in target systems
with moderate to small per CPU-core memory availability and/or when the input
data is large.

In order to address such problems, we proposed a new reuse approach called
Runtime Memory-Efficient Scheduler for Reuse (RMSR) that combines an off-line
analysis with an on-line scheduler to orchestrate the execution and maximize
reuse by removing the memory limitation aspect of RTMA. As a consequence, RMSR
improved RTMA in up to 2.8$\times$ in our experimental analysis.  We have also
executed large-scale runs on a distributed memory machine with 256 nodes (7,168
CPU cores) in which a parallel efficiency of about 0.92 was attained. This
demonstrates that the gains with the proposed optimizations are maintained in
large-scale SA runs. This level of performance improvement opens new
opportunities for using SA in our application domain.

The rest of this paper is organized as follows: Section~\ref{sec:rt-framework}
describes the Region Templates Framework (RTF) in which the proposed
optimization was build. RMSR is presented in Section~\ref{sec:rmsr}, and
experimental evaluation is detailed in Section~\ref{sec:results}. Further,
Sections~\ref{sec:related} and~\ref{sec:conclusions} discuss, respectively, the
related work and conclusions.

\section{Region Templates Framework}
\label{sec:rt-framework}

The Region Templates Framework (RTF) was developed to enable the execution of
dataflow applications on large scale distributed
environments~\cite{Teodoro2014589}. The main domains of the RTF include
data-intensive applications that use data elements represented in low
dimensional spaces (1D, 2D or 3D) with an optional temporal component. Examples
of these applications include medical image segmentation and object
annotation~\cite{kong2013machine}.

The RTF is comprised by the data abstraction model, the hierarchical storage
layer, and the runtime system. The data abstraction defines data object
structures that are commonly used. These objects are managed
by the hierarchical storage that saves and retrieves them in a
distributed memory machine.  The storage is optimized by using multiple memory
layers, such as RAM, SSD, HDD etc, that can be local and/or distributed in a
set of nodes.

The runtime system of the RTF allows for the application to be represented and
executed as hierarchical workflows, comprised of coarse-grain stages in which
each stage can be a workflow of fine-grain tasks. The execution on a
distributed memory machine follows a Manager-Worker model and stage instances
are assigned for execution with Workers in a demand-driven fashion. As such,
the fine-grain tasks that implement a given stage instance are executed within
a single Worker. Further, a Worker may use all computing resource available in
a node (CPU cores, GPUs, etc) by dispatching tasks in multiple devices
concurrently~\cite{Teodoro-IPDPS2013,Teodoro-IPDPS2012,cluster09george,doi:10.1177/1094342015594519}.

In the RTF, application stages communicate by writing/reading data objects from
the hierarchical storage. This feature (i)~alleviates the
application development effort as inter-node communication is managed by the
runtime; and, (ii)~also allows for the system to automatically perform
decisions about the data and execution placements with the goal, for
instance, of minimizing data movements.

\subsection{Executing Sensitivity Analysis (SA) in the RTF}
\label{sec:sa}

\begin{figure}[t]  
\begin{center}
\includegraphics[width=0.48\textwidth]{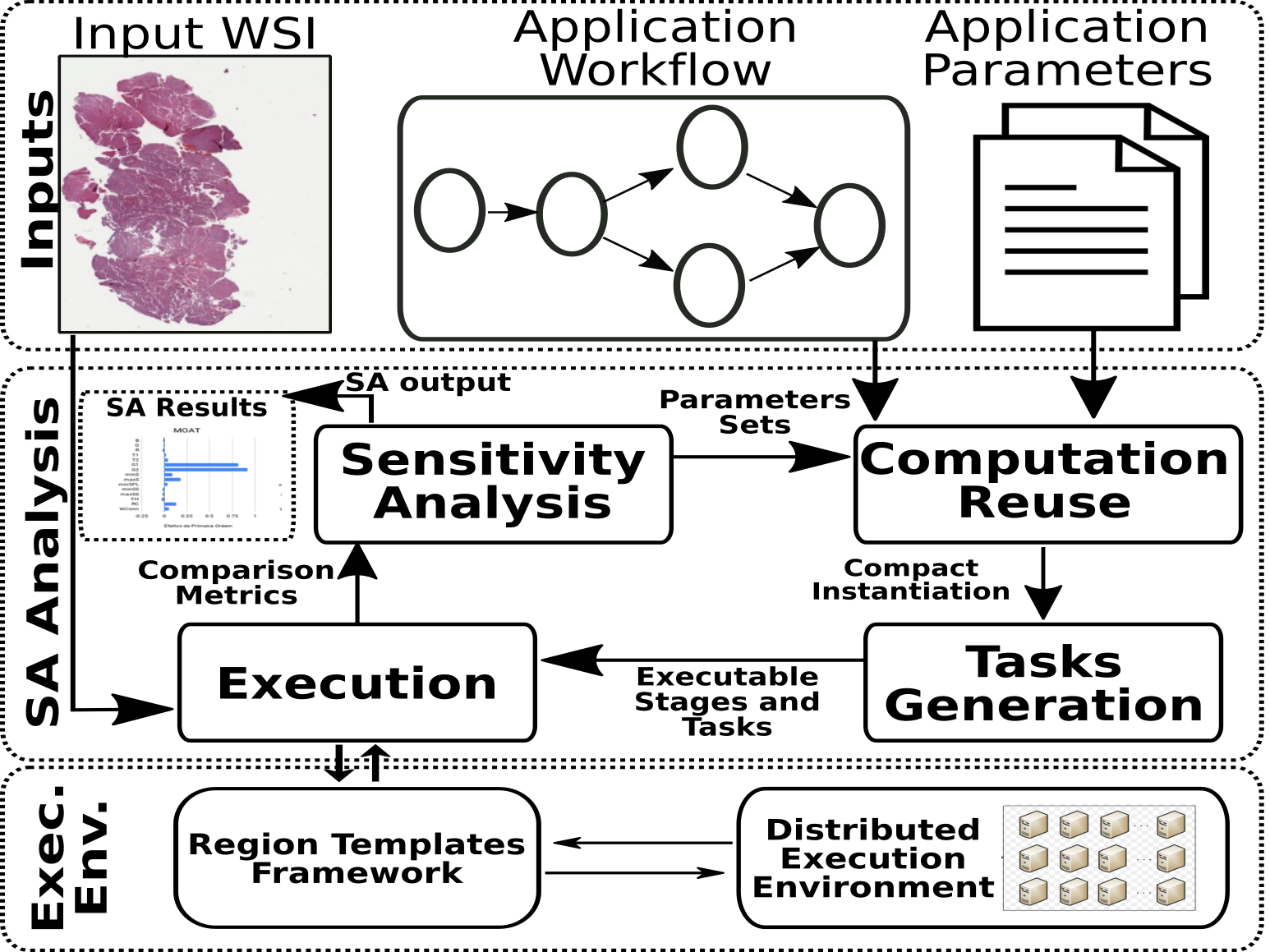} 
\vspace*{-4ex}
	\caption{Architecture of the overall SA framework.}
\vspace*{-4ex}
\label{fig:overview} 
\end{center} 
\end{figure}

This section describes the components built on top of the RTF for executing SA
(Figure~\ref{fig:overview}). A SA study receives the input data, application
workflow, and parameters to be analyzed. It then executes a SA method that will
select parameters' values sets for which the application should be executed.
Those parameters sets and application workflow are analyzed for computation
reuse, and the resulting workflow is dispatched for execution with the RTF in a
high-performance machine.  The result of this execution is a metric of
difference (e.g., Dice or Jaccard) that measures the difference between
application results (segmentation) for each parameter set vs. segmentation
results computed using the application default parameters. These metric values
are returned to the SA module that computes indices with the importance of
each parameter to changes in the output.

SA can be computed through a number of methods. Some include screening methods
as the Morris One-At-A-Time (MOAT)~\cite{10.2307/1269043}, which are commonly
employed to efficiently identify non-important parameters. Further, we can also
employ other more comprehensive methods that calculate importance measures,
such as Pearson's and Spearman's correlation
coefficients~\cite{saltelli2004sensitivity} or the Variance-based Decomposition
(VBD)~\cite{Weirs2012157}. These methods were listed according to their demands
in the number of application runs (sampling size), and they are typically used
in coordination. For instance, MOAT can be employed first to reduce the
parameters to a core set of known important ones, before other costly methods
are executed. The methods may use different approaches to select the parameters
sets to be evaluated: Monte Carlo sampling, Latin hypercube
sampling~(LHS)~\cite{10.2307/1268522}, quasi-Monte Carlo sampling with Halton
or Hammersley sequences, etc.

\subsection{Multi-level Computation Reuse}
\label{sec:reuse}

\begin{figure}[htb!]
\begin{center}
        \includegraphics[width=0.47\textwidth]{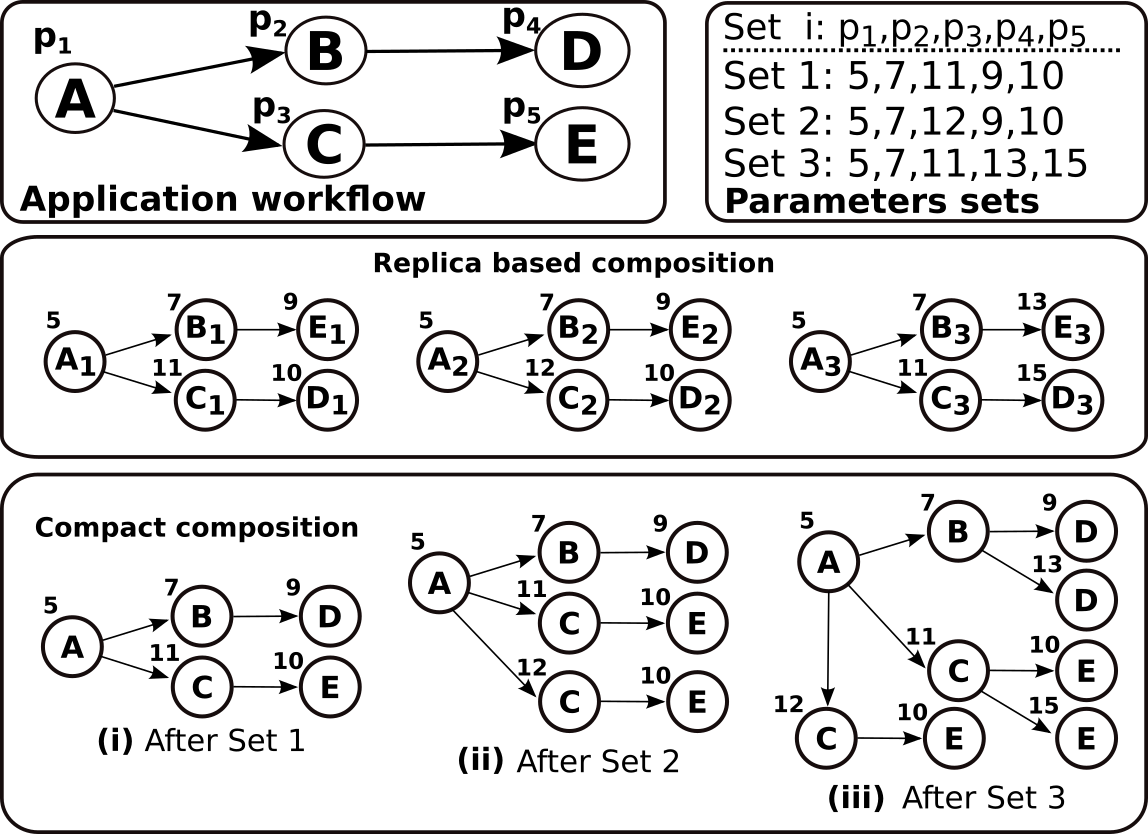}
\vspace*{-1ex}
\caption{Example of a workflow composition. This composition can be performed by either fully replicating the
base workflow for each parameters set, or performing a compact composition, which 
excludes the execution of repeated tasks.}
\vspace*{-3ex}
\label{fig:app-graph}
\end{center}
\end{figure}

The RTF used in this work performs the execution using hierarchical workflows
with two levels: stages and tasks.  This raises possibilities for reuse of
stages (coarse-grain) and tasks (fine-grain). In stage-level reuse the sets of
parameters' values for a given stage are compared to find stages that use
exactly the same input parameters. In that case, a single copy of a duplicate
stage is instantiated and dependencies in the workflow are fixed
so that downstream stages can receive the information. Figure~\ref{fig:app-graph} presents an example of an
application workflow and parameters sets to execute in a SA.  Further, it
depicts the application instantiation in the replica-based with no reuse and
the compact composition in which replicated stages are not executed
multiple times.  

Stage-level reuse can not take advantage of partial reuse among stages. For
that sake, task-level reuse has been proposed~\cite{8048914}. The task-level
reuse will merge together stage instances with overlapping (but not equal) set
of parameter values. The replicated tasks from the merged stage instances are
then removed, and the remaining ones will be part of a single, coarser-grain,
stage instance. This creates an additional challenging with respect to the
number of stages that can be merged. As the number of stages merged increases,
the amount of memory required to executed that stage also grows. A previous
work has implemented an algorithm called Reuse-Tree Merging Algorithm
(RTMA)~\cite{8048914} in which the maximum number of stages (MaxBucketSize) to
be merged is fixed and selected by the user.

\begin{figure}[htb!]
\begin{center}
        \includegraphics[width=0.47\textwidth]{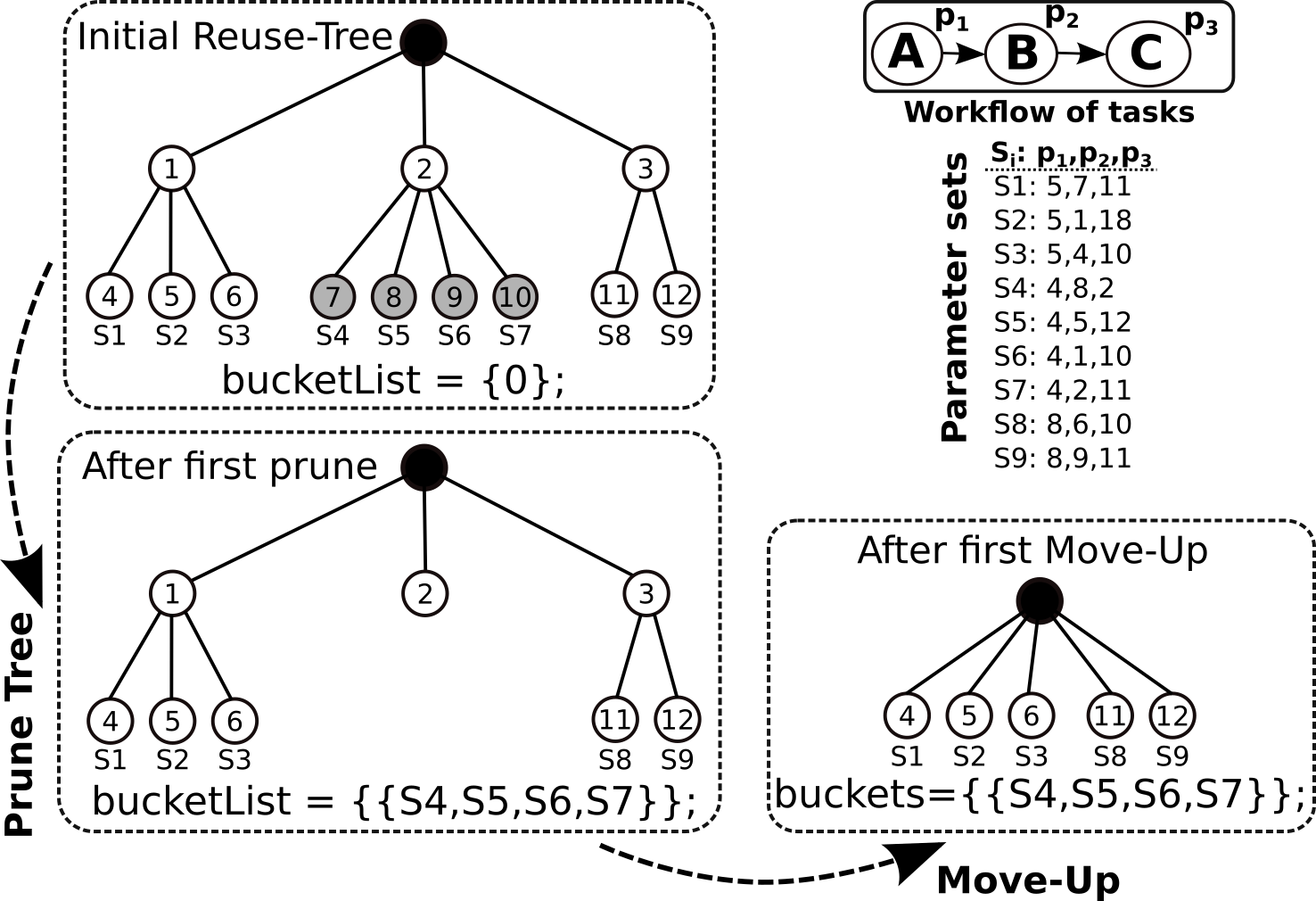}
\vspace*{-1ex}
\caption{RTMA example with 12 stage instances that are grouped in buckets of MaxBucketSize$=$4.}
\vspace*{-3ex}
\label{fig:reuse-tree-example}
\end{center}
\end{figure}

The RTMA organizes the stage's tasks in a tree structure where each task is a
node, as illustrated in Figure~\ref{fig:reuse-tree-example}. This example uses
a stage workflow with three tasks. Stages are assigned to different branches of
the tree according to the parameters' values used by their internal tasks.
After the tree is built, the deeper the first common ancestor is for a pair of
stages instances, the higher the amount of reuse among them. Due to the limit
in the number of stages that can be merged, RTMA generates buckets of
stage instances to be merged respecting the MaxBucketSize value.

The merging or choice of these buckets of stage instances is an iterative
process. The algorithm searches for MaxBucketSize stages with the same parent
in the tree that to fill a bucket, creates a bucket for
those stages, and removes them from the tree. This is computed for all leaf
nodes and corresponds to the prune phase of the algorithm. Further, the
remaining leaf nodes (stage instances not assigned to a bucket) are moved one
level up in the tree, and the same process is repeated until all stage
instances are assigned to a bucket.  An example of these operations for
MaxBucketSize=4 is presented in Figure~\ref{fig:reuse-tree-example}. Stages
S4-S7 have the same parent and are sufficient to fill a bucket, so they are
removed from the tree. Since other leaf nodes with a common parent are not
sufficient to create more buckets, the algorithm will go to the move-up phase.
Nodes 4-6 are moved to the parent node of their parents, as are nodes 11 and
12. These leaf nodes' parents are removed from the tree along with any other
childless node (e.g., node 2). Finally, the new tree can have the same process
re-executed until all stages are not assigned to a bucket.

\section{Runtime Memory-Efficient Scheduler for Reuse(RMSR)}
\label{sec:rmsr}

\begin{figure}
\begin{center}
		\includegraphics[width=0.95\columnwidth]{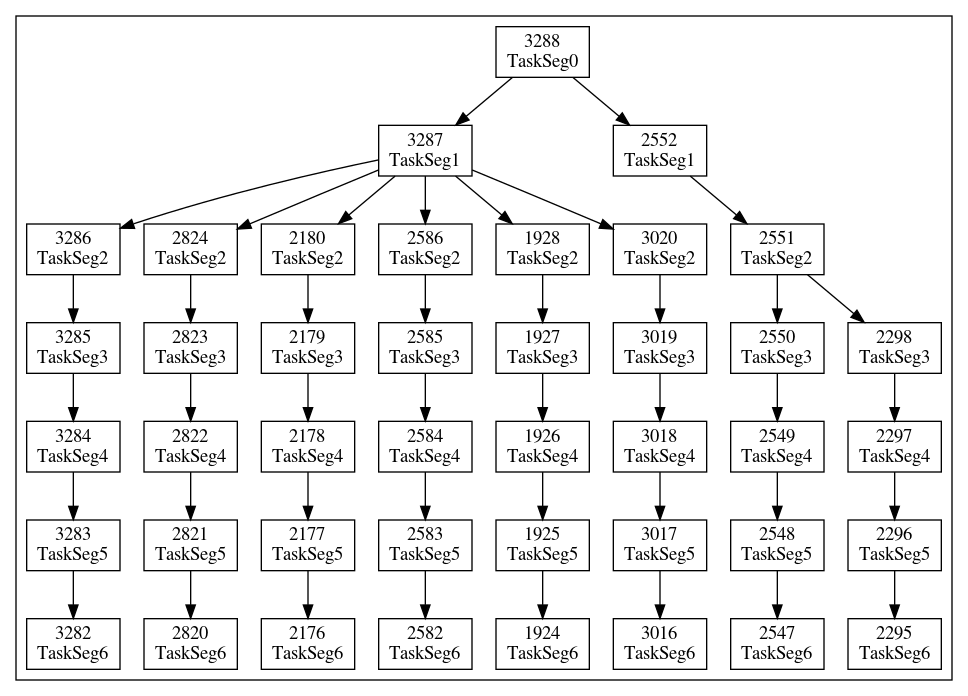}
	\caption{Typical workflow with 7 tasks (Seg0,...,Seg6) after passing a computation reuse analysis with a MaxBucketSize of 8: the resulting tasks tree of the merged stage has a width and the memory demands proportional to MaxBucketSize.}
\label{fig:rt-a}
\end{center}
\end{figure}

The main purpose of the RMSR algorithm is to address the suboptimal gains with
reuse in RTMA due to its limited merging (MaxBucketSize). 
A merged stage in RTMA will have a typical form of a tree
(Figure~\ref{fig:rt-a}), where the tree width is proportional to MaxBucketSize.
Because the memory is limited, the number of stages that can be merged
(MaxBucketSize) must be adjusted according.
One could think of using a disk storage as an auxiliary memory to allow for
larger MaxBucketSize and improve reuse. However, this is prohibitive in our use
case application because the tasks are rather fine-grain and storing results
outputted by all tasks (large images) in each stage is more costly than the
execution time of the tasks. 

The main idea of the RMSR is to modify the order in which tasks within a
coarse-grain stage are executed to decouple the MaxBucketSize from the memory
actually used to execute the application. It works as an additional run-time
phase to RTMA that receives a set of merged stages and limits the number of
branches of the task tree (Figure~\ref{fig:rt-a}) that are being
executed/active at any time in the execution.
This is done by allowing only a parameterized number of paths (activePaths)
from the root to the leafs of the tree of tasks representing a stage to be
concurrently active. 
For each of the active paths executed by RMSR, the tasks in that subtree are
processed in a depth-first order, which limits the number of processing paths
active (and using memory) regardless of the MaxBucketSize employed to build that
stage. As a consequence, arbitrary high MaxBucketSize values can be used to merge stages as long
as the activePaths are controlled during the execute to limit memory utilization.

\begin{algorithm}[htb]
\caption{RMSR tasks scheduler}\label{alg:rmsr}

\begin{algorithmic}[1]
    \State \textbf{Input:} $stageTaskTree$, $activePaths$
    \State $taskStack \leftarrow stageTaskTree.root$
    \While {$taskStask \neq \emptyset $ \textbf{or} $stageTaskTree \neq \emptyset $} 
	\If{$taskStask \neq \emptyset$ \textbf{and} $activePaths > 0$}
		\State $activePaths--$
            	\State $task \leftarrow taskStack.pop()$
		\State $task.run()$ 
		\State \Comment{Resolve deps/insert new tasks in the stack}
		\ForAll{$dep \in task.dependents$}
            	    \State $dep.nDependecies--$
                	\If{$dep.nDependecies == 0$}
	                    \State $taskStack.push(dep)$ 
			    \State $stageTaskTree.remove(dep)$
        	        \EndIf
	            \EndFor
		\State $activePaths++$
	\EndIf	
    \EndWhile
\end{algorithmic}
\end{algorithm}

The Algorithm~\ref{alg:rmsr} describes the main steps of the depth-first style
execution of a stage workflow of tasks. It receives as input the stage tree of
tasks and the number of active paths to use. The main loop from lines~3-14 will
execute until all tasks have been processed. It will then check whether there
are tasks available for execution (taskStack not empty) and the maximum number
of active paths was not reached. If both are true, a new task is selected for
execution from the top of the $taskStack$ to assert a depth order. The
task is executed (line~7) and tasks that depend on it are pushed to the
taskStack once they have all dependencies resolved. 

Although Algorithm~\ref{alg:rmsr} is presented in a sequential fashion, it is
executed by multiple threads in a Worker to select the next task to process.
Each thread will deal with a path per time.  Also, we want to highlight that
the algorithm developed was designed for applications with tasks that are
homogeneous in terms of memory demands. For heterogeneous cases, RMSR would
have to limit the number of active paths to that of the memory spent by the
most demanding tasks, which would be suboptimal.

-real
\section{Experimental Evaluation} \label{sec:results}

The experiments were executed using brain tissue images from cancer
studies~\cite{kong2013machine}. The application used in our evaluation consists
of the normalization, segmentation and comparison stages (see Figure
\ref{fig:example-app}). The comparison metric implemented is the Dice
coefficient of objects found for each parameter set used in the SA vs. those
generated using the application default parameters. The experiments were
executed on a cluster machine on which each node had a dual socket Intel
Haswell (E5-2695 v3) CPUs (14 cores each CPU), 32~GB of memory, and Red Hat
Linux. The machines are inter-connected with Infiniband switches, and codes
were compiled with Intel Compiler 13.1 using ``-O3''. All execution times refer
to the application makespan, including I/O, scheduling, and actual processing
times.

\subsection{The Impact of Computation Reuse to Performance}

This section presents the impact of performing computation reuse for SA runs.
In this setting, we have used a MOAT study with a two parameter sampling sizes
(number of application runs) created using a quasi-Monte Carlo with a Halton
sequence. We have compared the execution without reuse (No reuse), with reuse
at the Stage Level only, and with reuse at task level that employs the RTMA
proposed in the previous work. 

\begin{figure}[h]
\begin{center}
	\includegraphics[width=0.49\textwidth]{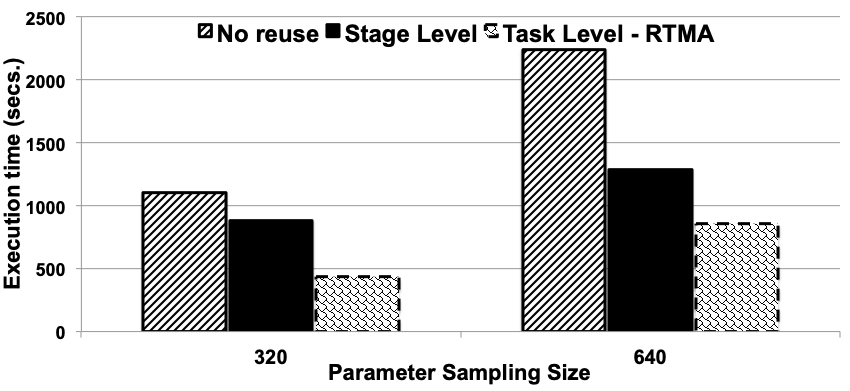}
\vspace*{-4ex}
	\caption{The performance benefits of different computation reuse strategies.}
\vspace*{-2ex}
	\label{fig:reuse-overall}
\end{center}
\end{figure}

The experimental results are presented in Figure~\ref{fig:reuse-overall}. As
shown, the gains with reuse of computation are significant for both strategies
and parameter sample sizes. In the case with 640 parameter sets, the Stage
Level reuse attained a speedup of about 1.7$\times$ as compared to not reusing
computation (No reuse). Further, the use of the task level reuse with RTMA
(multi-level reuse) resulted in gains of 2.6$\times$ and 1.5$\times$,
respectively, on top of the No reuse and Stage Level approaches. The experiment
also shows that gains with the reuse are higher with a larger parameter
sampling size, which was expected due to the higher probability of duplicate
computations in coarse(large images)-/fine-grain cases. While this section demonstrates that
the performance benefits with computation reuse can be very significant, the
actual performance attained by the RTMA is limited by number of stages it can
merge together (MaxBucketSize). In the next section we compare the performance
of RMSR designed to deal with the memory limitations to RTMA.

\subsection{The Impact of the Memory Availability to Computation Reuse (RMSR vs RTMA)}

This section compares the performance of RMSR to RTMA~\cite{8048914}. It is
performed under scenarios with different memory availability, which is the main
aspect affecting the RTMA performance. Also, we use a MOAT SA study with 800
parameter sets and input images with 4K$\times$4K pixels.

To limit the algorithms memory utilization, we have varied their parameters
with respect to bucket size and active paths used. The algorithms
configurations are defined as RTMA(Y,X) and RMSR(Y,X), where Y corresponds to
the internal parallelism (number of threads used in a Worker) and active paths
in the case of RMSR. The X parameter is the MaxBucketSize used. It is worth
recalling that RMSR is able to increase the X value while limiting the memory
utilization with the number of active paths. Each experiment employs a
fixed Y value in RTMA and RMSR to isolate the parallelism gains and focus on
the computation reuse. The Y or internal parallelism used is equals to the X
employed in RTMA.

\begin{figure}
        \centering
	\subfloat[RTMA(2,2) or no reuse vs. RMSR with varying bucket size - Memory availability of 6~GB.\label{fig:rmsr-1}]{
	   	\includegraphics[width=0.975\columnwidth]{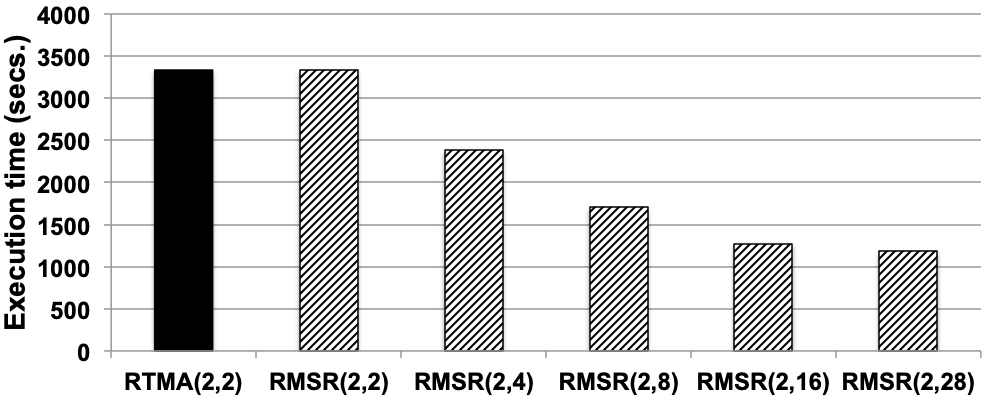}
	}\hfill
	\subfloat[RTMA(4,4) vs. RMSR with varying bucket size - Memory availability of 12~GB.\label{fig:rmsr-8}]{                		
		\includegraphics[width=0.975\columnwidth]{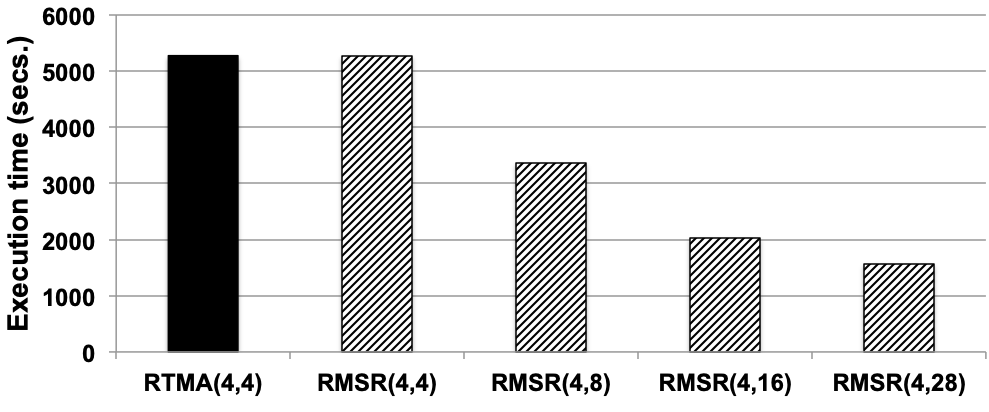}
	}\hfill
	\subfloat[RTMA(8,8) vs. RMSR with varying bucket size - Memory availability of 24~GB.\label{fig:rmsr-8}]{                		
		\includegraphics[width=0.975\columnwidth]{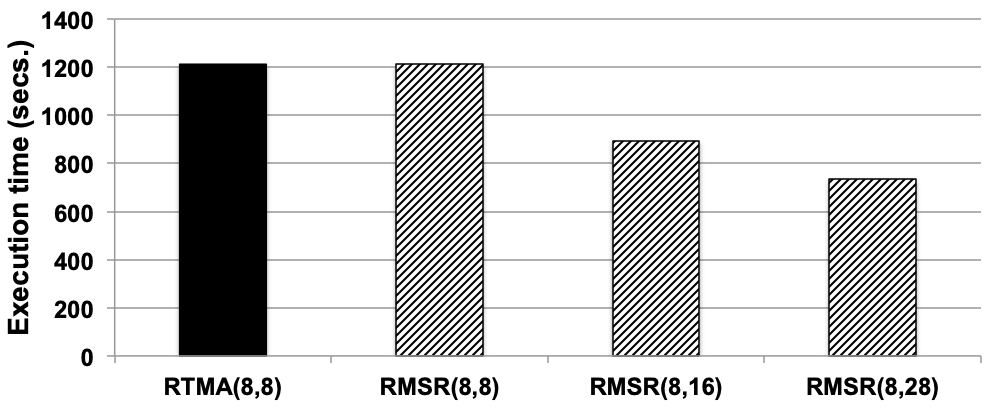}
	}\hfill
	\caption{Comparison of the RTMA and RMSR in settings with different memory availability. While the RTMA has to maintain a fixed upper bound for the computation reuse stage merging (MaxBucketSize) for each memory configuration, the RMSR is able to increase the merging size a control the memory use through the execution.\label{fig:rsmr-rtma}}
\end{figure}

To setup the maximum memory used in each experiment we used the minimum reuse
case (RTMA with MaxBucketSize of 2) as a baseline (6~GB), and have doubled the
memory availability until the RTMA execution fits into the 32~GB available in
our target machine. This results in the configurations with 6~GB, 12~GB, and
24~GB as presented in Figure~\ref{fig:rsmr-rtma}. First, for all memory 
configurations, the performance differences of RTMA and RMSR are negligible
when the same Y and X values are used. This indicates that the RMSR scheduling
costs are not significant. Further, when X (or MaxBucketSize) is increased for
RMSR while keeping Y (active paths and threads) the same as that used by RTMA,
the RMSR performance is improved. For instance, for (large images)the case with 6~GB of
memory (Figure~\ref{fig:rmsr-1}) RMSR(2,28) is 2.8$\times$ faster than RTMA
(2,2).  This performance gain tends to decrease when the memory availability is
higher, because RTMA can perform more reuse.  However, the RMSR
gains are still significant in the case with 24~GB of memory, since the
RMSR(8,28) is 1.6$\times$ faster than RTMA(8,8). In all cases, the gains are a
result of the RMSR ability to perform more aggressive merging (or computation
reuse) while restricting the memory utilization during the execution. 

\subsection{Impact of the Input Data Size to the Computation Reuse (RTMA vs RMSR)}

The input data size is another aspect that affects the computation reuse in
RTMA. As the data size used grows, the application workflow demands more memory
for execution, which results in the need of using smaller
MaxBucketSizes in RTMA. In this section, we quantify the amount of task reuse
that RTMA and RMSR can achieve when processing larger images (9K$\times$9K,
10K$\times$10K, and 11K$\times$11K pixels). We want to highlight that these
image sizes are a common case in our domain, because current microscopes can
generate images with resolutions in the order of 120K$\times$120K pixels.  For
that sake, we have measured the maximum bucket size (MaxBucketSize) that RTMA
would be able to use for potential target machines with larger memory: 64~GB
and 128~GB. Than, we computed the task reuse that RTMA and RMSR would be able
to perform for a VBD SA study with 8,000 parameter sets (runs). 

\begin{table}[h!]
\caption{Reuse attained by RTMA and RMSR for different image sizes and two potential target machines with 64~GB and 128~GB of memory.}
\label{tab:mbsreuse64}
\begin{tabular}{llllll}
\hline
             		&  		& \multicolumn{2}{c}{Machine with 64~GB}	& \multicolumn{2}{c}{Machine with 128~GB}  \\ \cline{3-6} 
             		& Image Size 	& BucketSize	& Reuse 	& BucketSize	& Reuse  	\\ \hline 
\multirow{3}{*}{RTMA} & 9K$\times$9K    & 4 	   	& 31.75\%       & 8 	   	&36.32\%	\\ \cline{2-6}
                      & 10K$\times$10K  & 3  		& 27.73\%       & 6  		&33.57\%	\\ \cline{2-6}
                      & 11K$\times$11K  & 2  		& 21.82\%	& 5  		&27.94\%	\\ \hline
RMSR                  & 9K,10K,11K  	& 10	   	& 36.36\%       & 10	   	&36.36\%	\\ \hline
\end{tabular}
\end{table}

The results are presented in Table~\ref{tab:mbsreuse64}, where BucketSize
refers to the MaxBucketSize that could be employed in RTMA for each pair image
size and memory available. The RMSR used a BucketSize of 10 that leads to reuse
close to maximum in this particular parameter set. As may be noticed, even with
machines with large memory sizes, the reuse with RTMA would be significantly
reduced with the use of big images, whereas using RMSR would preserve such
performance gains. For instance, for the case with 64~GB and images with
11K$\times$11K, RMSR would be able to reuse about 1.6$\times$ more tasks than
RTMA.

\subsection{Multi-core and Multi-node Scalability}

\begin{figure}
        \centering
	\subfloat[Multi-core execution.\label{fig:multi-core}]{
		\includegraphics[width=0.975\columnwidth]{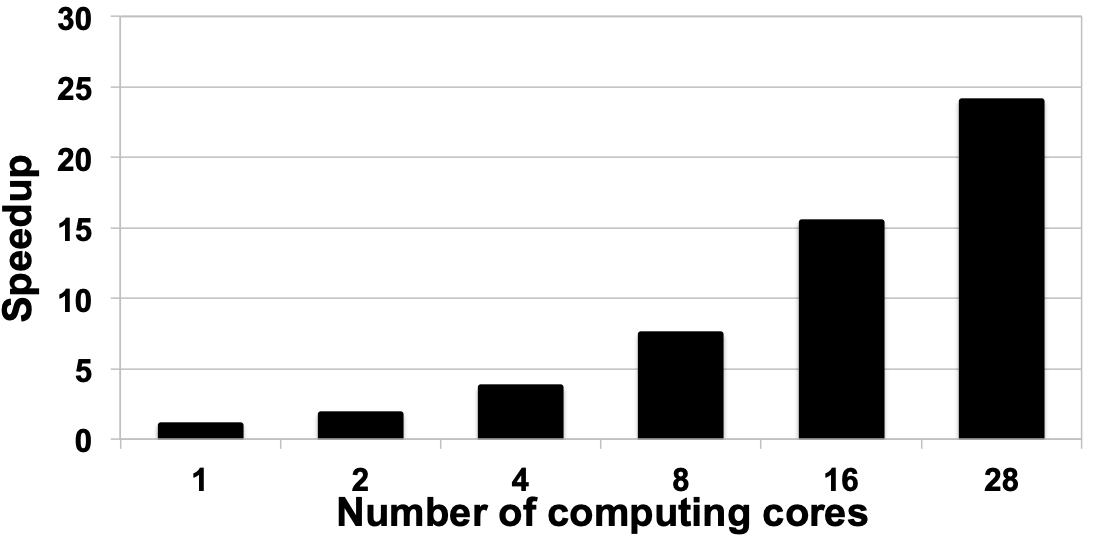}
	}\hfill
	\subfloat[Multi-node execution.\label{fig:multi-node}]{                
		\includegraphics[width=0.975\columnwidth]{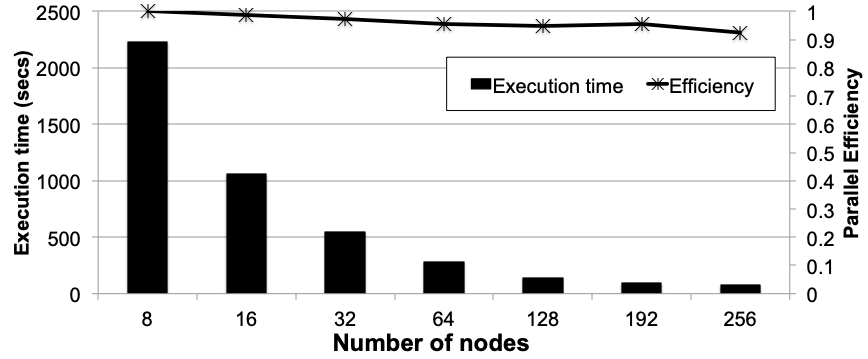}
	} 
        \caption{Execution scalability of our system in large-scale SA.\label{fig:scalability}}
\end{figure}

Finally, in this section, we evaluate the scalability of the execution with RMSR
in a multi-core and distributed memory environment. Figure~\ref{fig:multi-core}
shows that the application attains good speedups as the number of computing
cores increases. However, the speedups are slightly smaller than the ideal
speedup, which is expected in this case because the memory and overall I/O
subsystems are shared by the CPU cores (threads).

Further, we executed the application on a multi-node setting with a larger
imaging dataset of 6,113 4K$\times$4K brain tissue image tiles. The purpose of
this experiment is to evaluate the overall performance of our system and
optimizations on large-scale experiments. The execution times are presented in
Figure~\ref{fig:scalability}. As shown, the execution scaled well as the number
of machines used increases and it attained an parallel efficiency of about 92\%
for the configuration with 256 nodes. This demonstrates that the gains with our
optimizations are maintained in large-scale runs.

\section{Related Work} \label{sec:related}

Computation reuse has been extensively employed in multiple application
domains~\cite{10.2307/1269043,Sodani-1998,reuse5,reuse7,reuse8,reuse16,corr/abs-1903-05176}.
It can be categorized according to the reuse granularity (fine-grain,
coarse-grain or both), the reuse strategy (memoization or analytic), and
whether the implementation requires customized hardware. Fine-grain reuse can
lead to more reuse opportunities, but may also be more complex to exploit due to inherent
overheads. Also, the memoization that uses
caches in an attempt to attain reuse is more popular.  However, it is
inefficient in data-intensive applications due to the high caching amounts that
would be necessary. In this case, an analytic approach can be used to compute
reuse opportunities in an application and link the output of operations to the
locations in which it is to be reused, alleviating the caching pressure.

Other works employed reuse buffers to retrieve low-level instructions
sub-results~\cite{Sodani-1998}. Their solution either ends instruction
pipelines earlier, reducing resources conflicts, or breaks next instructions
dependencies through reuse, thus reducing the overall execution cost. Another
strategy on hardware~\cite{reuse4} has implemented an interesting reuse
approach with the goal of minimizing power consumption. In this case, they
would profile the application to quantify reuse regions, whose granularities
were chosen taking into account the benefits.

Computation reuse in distributed environments has been employed by caching
secure-function evaluation (SFE) that are used by multiple clients in a
server-client scheme~\cite{reuse5}. As one may notice, this strategy
would require a coarse-grain task reuse for attaining scalability. Computation
reuse has also been employed in bioinformatics~\cite{reuse7,reuse8}.
In~\cite{reuse7} outputs from full application runs are stored and reused,
whereas~\cite{reuse8} utilizes memoization to save partial results in the
process of comparing proteins.

More recently Riera et al.~\cite{reuse16} proposed using fine-grain reuse on
deep neural networks as a way to reduce energy consumption. They leveraged the
fact that such networks are tolerant to small precision variations, which
increases the amount of reuse opportunities.
This concept was implemented on hardware-level, but even if used as an
analytic method on software-level, its reliance on error-tolerant applications
makes it not applicable to the domain discussed in this work. Another recent
work by Li Liam et al.~\cite{corr/abs-1903-05176} has performed computation
reuse in application tuning. Their work represents the application as a directed
acyclic graph in which coarse-grain stages can be reused. Reuse is attained at
runtime by finding pre-computed results.  Still, the use of only coarse-grain
task with a memoization strategy limits its use with applications that deal with
small-scale datasets.

While there is a large number of works on computation reuse, we have not been
able to find a work that would lead to efficient reuse in our domain. Our
application is very compute intensive and manipulates large images, as such, requires execution on
high-performance distributed memory machines. The large datasets used and
processed are a limitation for using memoization, which is employed in must of
the literature. To overcome this limitation we developed a novel analytic
approach to identify reuse in the application workflow. Also, we have not seen
multi-level reuse in other works in the literature, neither have we found
approaches that leverage run-time scheduling to reduce the memory utilization
with the goal of improving reuse.

\section{Conclusions} \label{sec:conclusions}

Sensitivity analysis is an important tool employed on several domains, which
has its use in practice limited in pathology image analysis because of the high
computation costs involved. The computation reuse is one of the potential
optimizations that may reduce SA costs. A previous work~\cite{8048914} has
developed RTMA for that sake. However, the reuse capabilities of RTMA are
limited by the memory available in the target machine, which would
significantly reduce the optimization opportunities in several scenarios. To
overcome this limitation, we proposed and implemented RMSR in this work. RMSR
allows for aggressive computation reuse without increasing the memory demands
as in the case of RTMA.  As a consequence, RMSR has been able to improve
the performance of RTMA in up to about 2.8$\times$. Further, we have also shown
that the gains of RMSR were maintained in large-scale runs in which a parallel
efficiency of about 0.92 was attained in a distribute memory machine with 256
nodes. This level of performance should allow SA to be more widely use in the
pathology image analysis.

As a future work, we intend to evaluate other optimizations to improve the
performance in computation reuse. For instance, we want to analyze the impact
of automatically adjusting the tasks granularity with the goal of maximizing
reuse. It will be interesting to see the impact of the workflow generated in that case
with respect to the tasks granularity and the potential effects in scheduling
overheads, for instance, if tasks are too fine-grained. We also argue that a single
workflow generation is suboptimal, and that the best approach would also
consider the input set of parameters to be evaluate with the application. In
this case, the application execution graph can be enriched with the actual
reuse information derived from parameters.

\noindent {\bf Acknowledgments}.
This work was supported in part by 1U24CA180924-01A1 from the NCI,
R01LM011119-01 and R01LM009239 from the NLM, CNPq, Capes/Brazil grant
PROCAD-183794, and NIH K25CA181503. This research used resources of the XSEDE
Science Gateways program under grant TG-ASC130023.

%


\newcommand{\BIBdecl}{\setlength{\itemsep}{0.2 em}}
\linespread{0.85}
\bibliographystyle{IEEEtran}
\bibliography{george}

\end{document}